\documentclass[aps,preprint,showpacs]{revtex4}

\usepackage{graphicx}
\usepackage{color}
\usepackage{amsbsy,amsmath}
\usepackage{subfigure}
\usepackage{bm}

\newcommand{\bs}[1]{\boldsymbol{#1}}
\newcommand{\be}{\begin{equation}}
\newcommand{\ee}{\end{equation}}

\newcommand{\modif}{}
\newcommand{\correc}{}

\begin{document}

\def\d{\mbox{\rm d}}

\title{Cosmology in One Dimension: Vlasov Dynamics}

\author{Giovanni Manfredi}
\email{manfredi@unistra.fr}
\affiliation{Institut de Physique et Chimie des Mat\'eriaux de
Strasbourg, CNRS and Universit\'e de Strasbourg, BP 43, F-67034 Strasbourg Cedex 2, France }
\author{Jean-Louis Rouet}
\affiliation{Universit\'e d'Orl\'eans, CNRS/INSU, BRGM, ISTO, UMR7327, F-45071 Orl\'eans, France.}
\author{Bruce Miller and Yui Shiozawa}
\affiliation{Department of Physics and Astronomy, Texas Christian University, Fort Worth, TX 76129}

\date{\today}

\begin{abstract}
Numerical simulations of self-gravitating systems are generally based on N-body codes, which solve the equations of motion of a large number of interacting particles. This approach suffers from poor statistical sampling in regions of low density. In contrast, Vlasov codes, by meshing the entire phase space, can reach higher accuracy irrespective of the density. Here, we performed one-dimensional Vlasov simulations of a long-standing cosmological problem, namely the fractal properties of an expanding Einstein-de Sitter universe in Newtonian gravity. The N-body results were confirmed for high-density regions and extended to regions of low matter density, where the N-body approach usually fails.
\end{abstract}



\maketitle

\section{Introduction}\label{sec:intro}
In the present epoch, the observable universe is extremely inhomogeneous. Taking galaxies for the essential elements, we see them grouped in clusters that are, in turn, grouped in superclusters, and interlaced with enormous voids. The recent detection of the local supercluster Laniakea \cite{Tully2014aa} exemplifies this scenario. Consideration of these gross features lead Mandelbrot \cite{mandel82} and Pietronero \cite{pietronero1987}, among others, to conjecture that the universe is a fractal, at least at some intermediate scales. An early thinker along these lines was de Vaucouleurs \cite{Vau}.  Since cosmological theory demands that the universe is homogenous at sufficiently large scales, the search for the transition to homogeneity has been a focus of recent investigations \cite{hogg2005cosmic,bagla2008fractal,yadav2010fractal,scrimgeour2012wigglez,labini2014spatial}.
Partial evidence for the fractal geometry is provided by the two-body correlation function computed from recent large-scale galaxy surveys like the Sloan or 2-degree \cite{Estrada:2009ApJ}, which exhibits power-law behavior over a finite range of scales. However, it is difficult to determine the mechanism and evolution of such scaling behavior from either observation or three-dimensional (3D) N-body simulations. To obtain a more complete picture, investigators have turned to lower-dimensional models where a more precise representation of the gravitational field is possible over all scales.

A great deal of work on structure formation in the universe has been accomplished using Newtonian 1D models (for a review see \cite{MR2010}, and for more recent work \cite{Joyce2011,Benhaiem11032013,MRS:2015arX}).
The link between 1D and 3D cosmology models was discussed in \cite{Benhaiem21092014}.
Nevertheless, N-body simulations, even in 1D, suffer from an intrinsic undersampling of the phase space because of the finite number of particles used in the codes \cite{Binney}. As in reality the number of bodies is virtually infinite, one should instead solve the mean-field Vlasov equation (which is the $N \to \infty$ limit of the N-body model) for the phase-space distribution coupled to the Poisson equation for the gravitational field. This is a very demanding computational task, both in terms of run duration and memory storage, particularly for situations where high accuracy is necessary to resolve the intricate phase space structures that develop over time. However, present computers now make this approach feasible, if not in 3D at least for 1D models.

{\correc
This work is devoted to the presentation of the first cosmological results obtained with a 1D Vlasov approach. The paper is organized as follows. In Sec. \ref{sec:model} we will introduce a set of scaled variables (comoving coordinates) that are particularly adapted to the Vlasov approach. Some details of the numerical algorithm used to solve the Vlasov-Poisson equations are provided in Sec. \ref{sec:numeric}. The numerical results are presented and analyzed in Sec. \ref{sec:results}. General conclusions are presented in Sec. \ref{sec:discussion}.
}

\section{Model and scaled variables}\label{sec:model}
Let us consider a highly symmetric expanding distribution of matter.
Its gravitational field has only one component $E_r(r,t)$ which depends on time and on a single spatial variable $r$. The symmetry could be, for instance, spherical or planar. In the planar case, originally developed by Rouet and Feix (RF) \cite{Rouet1,Rouet2} and later expanded by Miller and Rouet \cite{MRexp,MRGexp,miller2010ewald}, the system is constituted of many parallel expanding planar sheets whose surface density decreases following the expansion law.
For spherical symmetry (the so-called quintic model \cite{quintic}) the system is composed of concentric spherical shells.

The equation of motion of a particle in such a Newtonian gravitational field reads as
\be
\frac{d^2 r}{dt^2} = E_r(r,t),
\label{motion}
\ee
where $r(t)$ is a spatial position in the expanding universe.
In the mean-field limit, the gravitational field is a solution of the Poisson equation
\be
\nabla_r \cdot {\mathbf E}=-4\pi G\rho,
\label{poisson}
\ee
where $\rho(r,t)$ is the matter density.
In order to account for the expansion, we transform space and time according to:
\begin{eqnarray}
r &=& C(t) \xi, \label{scaler} \\
dt &=& A^2(t) d\theta,\label{scalet}
\end{eqnarray}
where $\xi$ is a comoving spatial coordinate. With this scaling, the velocity variable is transformed as
\be
v = CA^{-2} \eta + {\dot C} \xi,
\label{scalev}
\ee
where $v=dr/dt$ and $\eta=d\xi/d\theta$ (a dot denotes differentiation with respect to the time $t$). Here $A(t)$ and $C(t)$ are two strictly positive functions of time.
The general equation of motion in the scaled variables reads as
\be
\frac{d^2 \xi}{d \theta^2}+2\,A^2\,\left(\frac{\dot C}{C}-\frac{\dot A}{A}\right)\,\frac{d \xi}{d \theta}+A^4\frac{\ddot C}{C}\,\xi=\frac{A^4}{C^3}\, \mathcal{E}\,,
\label{general_scaling}
\ee
where $\mathcal{E}(\xi,\theta)= C^{2}(t)E_r(r,t)$ is the scaled gravitational field, satisfying
\be
\nabla_\xi \cdot \bs{\mathcal{E}}=-4\pi G\hat\rho,
\label{poissonscaled}
\ee
and $\hat\rho(\xi,\theta) = C^{3}(t)\rho(r,t)$
so that the total mass is preserved.

For the scaling functions, we  use the following forms:
\be
A^2(t)=(t/t_0)^\beta~;~~~~~
C(t)= (t/t_0)^{\gamma}
\label{ACgeneral}
\ee

\subsection{Standard scaling}
The standard scaling \cite{Rouet1,Rouet2} uses $\beta=1$ and $\gamma=2/3$. With this choice, all coefficients in Eq. (\ref{general_scaling}) become time-independent:
\be
\frac{d^2 \xi}{d\theta^2} + \frac{1}{3t_0}\frac{d \xi}{d\theta} -\frac{2}{9t_0^2}\xi= \mathcal{E}(\xi,\theta).
\label{scaledstandard}
\ee
For a constant density $\hat\rho = \rho_0$, Poisson's equation (\ref{poissonscaled}) can be solved exactly in a $d$-dimensional space
to give the gravitational field $\mathcal{E}=-4\pi G \rho_0 \xi/d = -\omega_J^2 \xi/d$, where $\omega_J = \sqrt{4\pi G \rho_0}$ is the Jeans frequency. At equilibrium, the gravitational field must exactly cancel the inverse harmonic term on the left-hand side of Eq. (\ref{scaledstandard}). This provides the relationship between $t_0$ and $\omega_J$:
\be
\omega_J^2 t_0^2 = {2\over 9}d.
\ee
Therefore, Eq. (\ref{ACgeneral}) can be written as:
\be
A^2(t) = (\alpha\,\omega_{J}t)^\beta ~;~~~~
C(t)= (\alpha\,\omega_{J}t)^{\gamma},
\ee
with $\alpha=1/(\omega_{J}t_0)=3/\sqrt{2d}$, so that the scaled and unscaled co-ordinates coincide at $t=t_0$.
Then, Eq. (\ref{scaledstandard}) becomes
\be
\frac{d^2 \xi}{d\theta^2} + \frac{\omega_J}{\sqrt{2d}}\frac{d \xi}{d\theta} -\frac{\omega_J^2}{d}\xi= \mathcal{E}.
\label{RF}
\ee
The RF model (considered here) has planar symmetry and is therefore essentially one-dimensional ($d=1$). The corresponding scaled Poisson equation is also 1D: $\partial_\xi \mathcal{E}=-4\pi G \hat\rho$.

For the quintic model \cite{quintic}, which corresponds to a spherically-symmetric expanding universe, we have $d=3$. If we consider a planar perturbation in the scaled universe, we can still use the 1D Poisson equation as above; however, the factor in front of the background term (third term on the left-hand side) of Eq. (\ref{RF}) must be modified as follows: $\frac{\omega_J^2}{d} \to \omega_J^2$, in order to allow for a steady state at equilibrium.
(Note that there is no such change in the RF model, because both the original system and the perturbation have planar symmetry).
With this substitution, the scaled equations of motion of the RF and quintic models only differ  in the coefficient of the friction term, and can be written as
\be
\frac{d^2 \xi}{d\theta^2} + \frac{1}{\sqrt{2d}}\frac{d \xi}{d\theta} -\xi= \mathcal{E}.
\label{RF2}
\ee
In the above equation we also introduced nondimensional variables, whereby the scaled time $\theta$ is normalized to the inverse Jeans frequency, the scaled space coordinate $\xi$ is normalized to an arbitrary length $\lambda$, and the scaled gravitational field $\mathcal{E}$ to $\lambda\omega_J^2$. We keep the same symbols for the nondimensional variables, which will be used throughout the rest of this work.
The  nondimensional Poisson equation reads as: $\partial_\xi \mathcal{E}=-\hat\rho$ where the density is normalized to $\rho_0$.

According to Eq. \eqref{RF2}, if the universe is homogeneous and strictly follows the expansion factor $C(t)$, then it will be static in the scaled variables, with a constant (nondimensional) density equal to unity and corresponding gravitational field $\mathcal{E}=- \xi$.
However, this is an unstable equilibrium which, when slightly perturbed, evolves towards a highly inhomogeneous distribution of matter with complex features.

\subsection{New scaling}
It is important to note that, of the two exponents $\beta$ and $\gamma$ in Eq. \eqref{ACgeneral}, only $\gamma$ has some physical bearing: it represents the rate of expansion of a self-similar Einstein-de Sitter universe. Instead, the exponent $\beta=1$ was chosen on purely utilitarian grounds to render the scaled equations autonomous. This is an appropriate choice for N-body simulations, because the 1D equations of motion can be solved exactly to machine precision, but a poor choice for a grid-based Vlasov code. Indeed, using this scaling, the transformed velocity $\eta$ grows exponentially in time, as was shown by N-body simulations (see Appendix \ref{app:newscaling}). This is a serious problem for Vlasov simulations, which solve the Vlasov equation on a fixed grid that covers the entire relevant phase space. If the velocities keep growing, one would need to mesh an increasingly large phase space in the $(\xi,\eta)$ variables, soon reaching memory and computation time limits.

The important point is that we can choose a different value of the exponent $\beta$ (and thus rescale time and velocity in a different way) so that the scaled phase space stays bounded for the entire duration of the run.
This can be achieved by choosing:
$\beta=(3\alpha-1)/(3\alpha) \approx 0.84$ in Eq. \eqref{scalet} (details of the calculations are given in the Appendix \ref{app:newscaling}). With this value, and using the same normalization as in Eq. (\ref{RF2}), one obtains the scaled equation of motion:
\be
\frac{d^2 \xi}{d \theta^2}+\frac{K}{\mu(\theta)}\,\,\frac{d \xi}{d \theta}-\frac{\xi}{\mu^2(\theta)}=\frac{\mathcal{ E}}{\mu^2(\theta)}\,,
\label{eqmotion_newscaling}
\ee
where $K=(3+\sqrt{2})/(3\sqrt{2})$, and $\mu(\theta) = \theta/3+1$.
Note that the scaled time $\theta$ depends on the value of $\beta$, which is not the same for Eq. (\ref{RF2}) (standard scaling, $\beta=1$) and for Eq. (\ref{eqmotion_newscaling}) (new scaling, $\beta \approx 0.84$).
Therefore, the two scaled times are not the same and their relationship is given in the Appendix \ref{app:time}.

The Vlasov-Poisson equations corresponding to Eq. \eqref{eqmotion_newscaling} reads as follows:
\begin{eqnarray}
\frac{\partial F}{\partial \theta} &+&
\eta\frac{\partial F}{\partial \xi} +  \frac{\partial}{\partial \eta} \left( \frac{\widetilde{\mathcal{E}}}{\mu^2(\theta)}~F - \frac{K}{\mu(\theta)}~\eta F  \right)
= 0, \label{vlasov_rescaled} \\
\frac{\partial \widetilde{\mathcal{E}}}{\partial \xi} &=& 1- \int_{-\infty}^{\infty} F(\xi,\eta,\theta)~ d \eta,
\label{poisson}
\end{eqnarray}
where $F(\xi,\eta,\theta)$ is the distribution function in the rescaled phase space. Note that, by defining $\widetilde{\mathcal{E}}=\mathcal{E}+ \xi$, the harmonic term in Eq. \eqref{eqmotion_newscaling} has been incorporated into Poisson's equation \eqref{poisson}.

\section{Numerical method and initial conditions}\label{sec:numeric}
We solved numerically the set of equations \eqref{vlasov_rescaled}-\eqref{poisson} with periodic boundary conditions in the scaled spatial variable $\xi$. Vlasov codes work by covering the entire phase space $(\xi,\eta)$ with a uniformly spaced grid. The distribution function $F$ is pushed in time using a split-operator scheme that treats the space and the velocity coordinates separately \cite{Cheng}.
{\modif
The time integration between $\theta$ and $\theta + \Delta \theta$ is performed in three steps. First we solve the equation
\be
\frac{\partial F}{\partial \theta} +
\eta\frac{\partial F}{\partial \xi}=0
\label{eq:step1}
\ee
whose exact solution is just a rigid shift of $\eta \Delta\theta$ in position space:
$F(\xi,\eta,\theta + \Delta \theta)=F(\xi-\eta \Delta \theta,\eta,\theta)$.
Then, the gravitational field $\widetilde{\mathcal{E}}$ is obtained through Poisson's equation \eqref{poisson}.
Finally, we solve the equation
\be
\frac{\partial F}{\partial \theta} +  \frac{\partial}{\partial \eta} \left( \frac{\widetilde{\mathcal{E}}}{\mu^2(\theta)}~F - \frac{K}{\mu(\theta)}~\eta F\right) =0 .
\label{eq:step2}
\ee
Because of the presence of the friction term and the time-dependent coefficients, this step is not standard. However, Eq. \eqref{eq:step2} can also be solved exactly (considering $\widetilde{\mathcal{E}}$ constant), by integrating the characteristic:
\be
\frac{d \eta}{d \theta} =
\frac{\widetilde{\mathcal{E}}}{\mu^2(\theta)} -\frac{K}{\mu(\theta)}\eta\,,
\ee
which has the following solution
\[
\Delta \eta\equiv \eta(\theta+\Delta\theta)-\eta(\theta)=C_\eta \eta(\theta) + C_\mathcal{E} \widetilde{\mathcal{E}},
\]
where $C_\eta = (B^{-3K}-1)$,
\[
C_\mathcal{E} = \left(B^{-3K}-\frac{1}{B} \right) \frac{1}{K-1/3}\, \frac{1}{1+\theta/3},
\]
and
\[
B = \frac{1+(\theta+\Delta\theta)/3}{1+\theta/3}.
\]
The solution of Eq. \eqref{eq:step2} is then:
$F(\xi,\eta,\theta + \Delta \theta)= F(\xi,\eta- \Delta\eta,\theta)$.
}

Interpolations on the phase-space grid are performed using an accurate finite-volume algorithm \cite{Filbet} that preserves the positivity of the distribution function. Vlasov codes display very low numerical noise even in regions where the matter density is rarefied, which is where N-body codes would be most noisy because of poor statistical sampling.
The Vlasov approach is widely employed in plasma physics, and has been used occasionally to study self-gravitating systems \cite{Mineau, Colombi_henon}, but was never applied to cosmological simulations.
A complementary approach to either N-body or Vlasov codes is provided by the water-bag method \cite{Colombi_wb}.

Equations \eqref{vlasov_rescaled}-\eqref{poisson} were solved for $0 \le \xi \le L$ and $-\eta_{max} \le \eta \le \eta_{max}$, with $L=10^4\pi$ and $\eta_{max}=15$. For the present results, we used $N_x=2^{15}$ points in the spatial coordinate and $N_v=1000$ points in velocity space.

The initial condition is a ``cold" Maxwellian in velocity space, with variance $\langle \eta^2 \rangle = 0.01$. The initial matter density $\rho(\xi)=\int F d\eta$ is so constructed as to display a power-law spectrum of the type: $P(k) \equiv |\rho_k|^2 \sim k^3$, where $k$ is the wave number in $\xi$-space.
{\correc
Initial power spectra of the form $P(k)\sim k^n$, with $n\in [0,4]$,
were used in a number of earlier works on structure formation \cite{Joyce2011,Benhaiem11032013,MRS:2015arX}.

The case $P(k)\sim k^3$ produces the 1D version of the Harrison-Zeldovich spectrum from the assumption of scale-free potential fluctuations \cite{peacock,Martinez-Saar}.
Following inflation, density fluctuations in the universe can be modeled as a Gaussian random field with a scale-free (power law) power spectrum $P(k)$. In a 3D universe, the exponent corresponding to a scale-free potential is unity, i.e., $P(k) \propto k$. To see this, we expand the gravitational potential ${\Phi}$ in a Fourier series \cite{peacock,Martinez-Saar}:
\begin{equation}
\Phi(\vec{r}, t) = V^{-1/2} \sum_{\vec{k}} \Phi_{\vec{k}} \exp(i \vec{k} \cdot \vec{r}),
\end{equation}
where $V$ is the volume.
Then it can be shown via the Poisson equation that
\begin{equation}\label{eq: potential_fluctuation_1}
\langle \Phi^2 \rangle = \frac{1}{2\pi^2} \int^{\infty}_0{k^2 \langle | \Phi_{\vec{k}} |^2 \rangle}dk = \int^{\infty}_0 \frac{P(k)}{k^2} dk=  \int^{\infty}_{0} \frac{P(k)}{k} d(\ln k).
\end{equation}
Consequently, if $P(k) \propto k$, the potential fluctuations are scale-invariant on a logarithmic scale. Initial conditions for 3D simulations of the expanding universe are guided by these considerations.

Similarly, in 1D we have:
\begin{equation}
\langle \Phi^2 \rangle = \int^{\infty}_0 dk \langle |\Phi_{\vec{k}}|^2 \rangle = \int_{0}^{\infty} \frac{P(k)}{k^3}d(\ln k).
\end{equation}
Requiring scale-free fluctuations for the potential then yields $P(k) \propto k^3$, which is the initial spectrum that we took for our 1D simulations.
}

In order to accelerate the early evolution, this initial condition was first allowed to evolve for a short time according to Eqs. \eqref{vlasov_rescaled}-\eqref{poisson} where the scale factor $\mu(\theta)$ was set equal to unity. Once the fluctuations have reached a sufficiently high level, the correct scaling was applied and the initial time was reset to $\theta=0$.

\begin{figure}[]
{\includegraphics[width=0.45\textwidth,height=12cm]{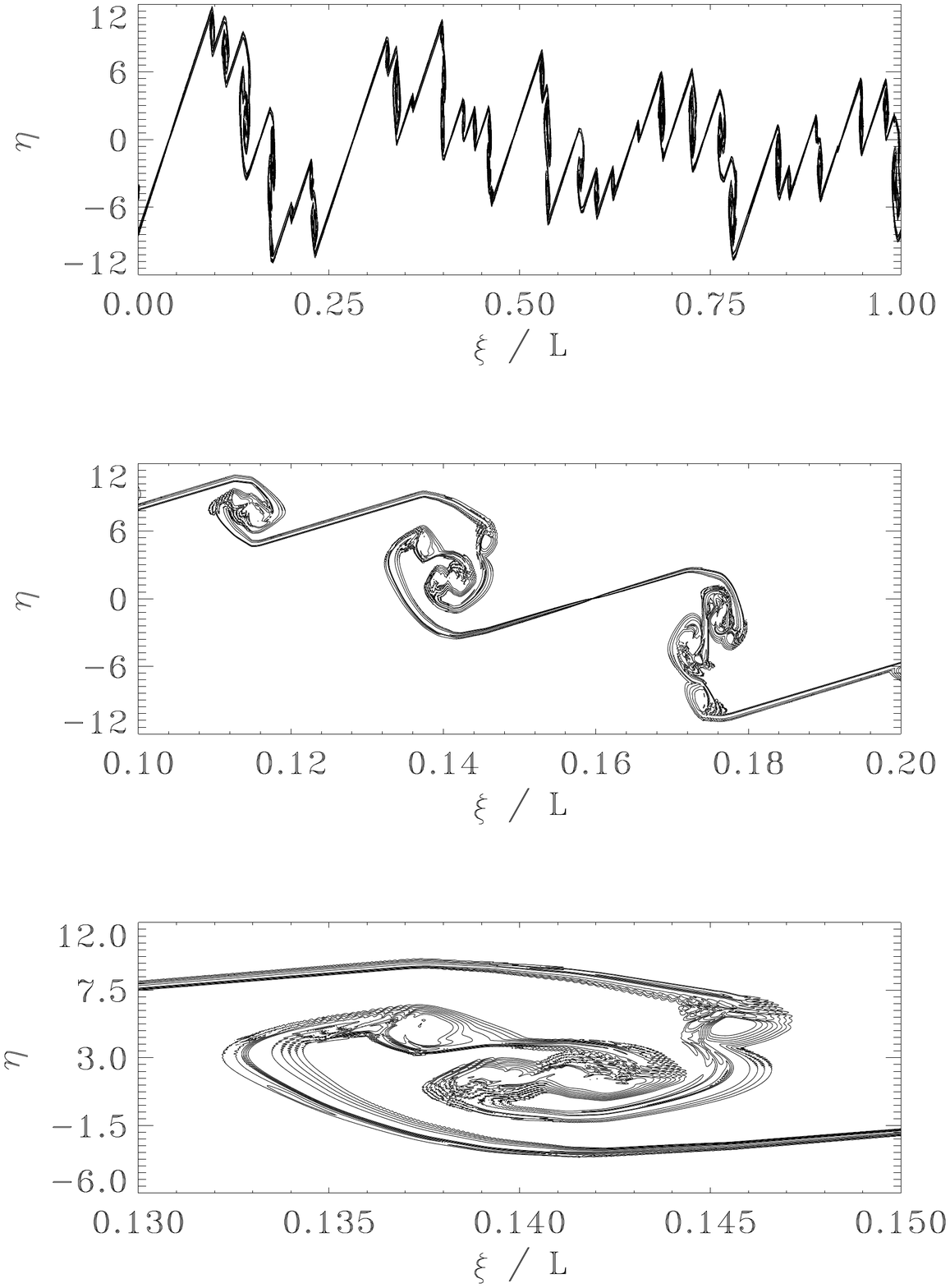}}
{\includegraphics[width=0.45\textwidth,height=12cm]{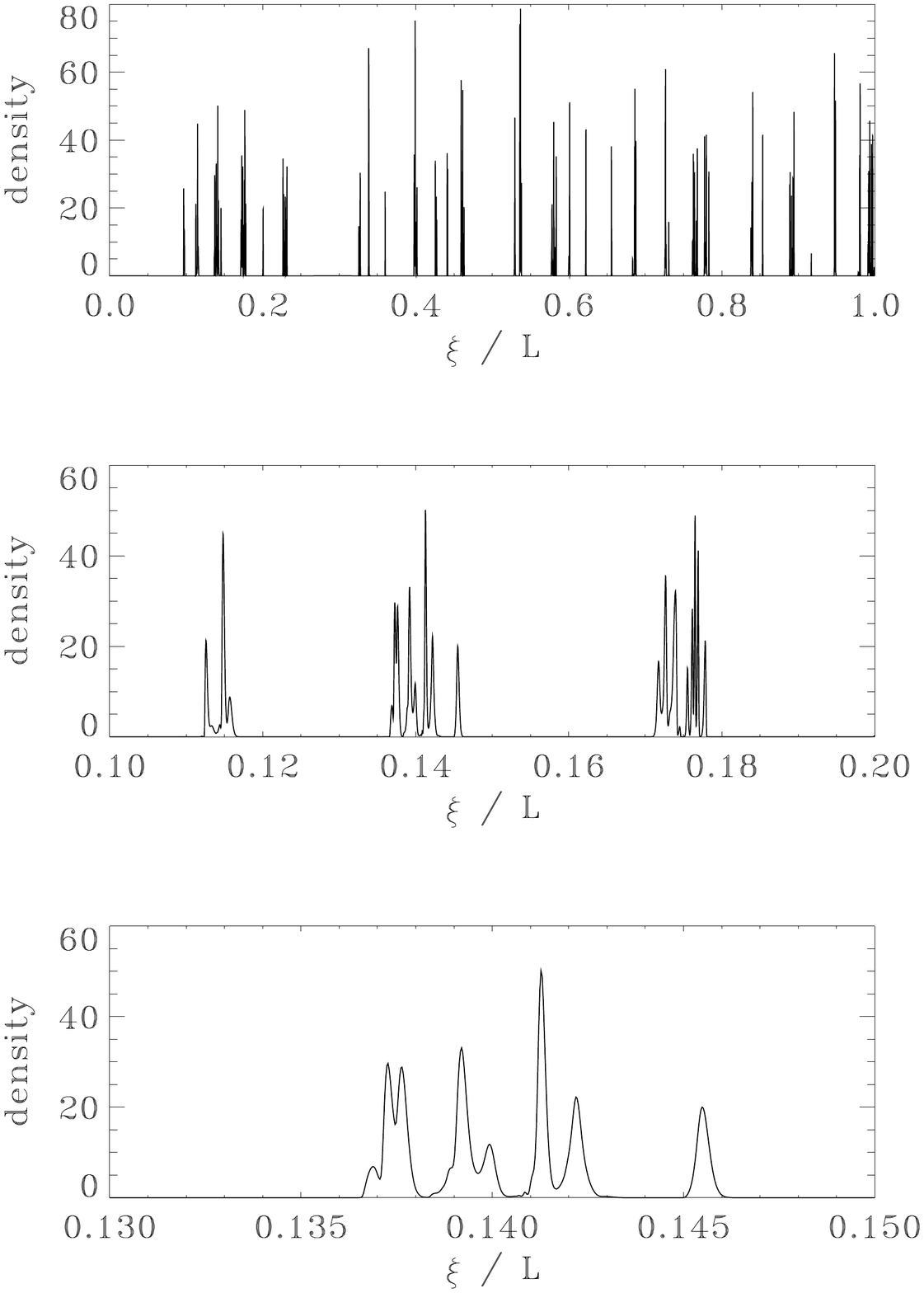}}
\caption{Phase-space distribution functions (left) and matter densities (right) at time $\theta=325$. The top panels show the entire domain, the center and bottom panels show consecutive zooms. The contour levels of the distribution functions are distributed logarithmically in the interval: $10^{-8} \le F \le 1$. The actual maximum value of $F$ is around $F_{max} = 185$.}
\label{fig:phasespace}
\end{figure}

\section{Results}\label{sec:results}
\subsection{Phase space and matter density}
\begin{figure}[]
{\includegraphics[scale=0.5,angle=90]{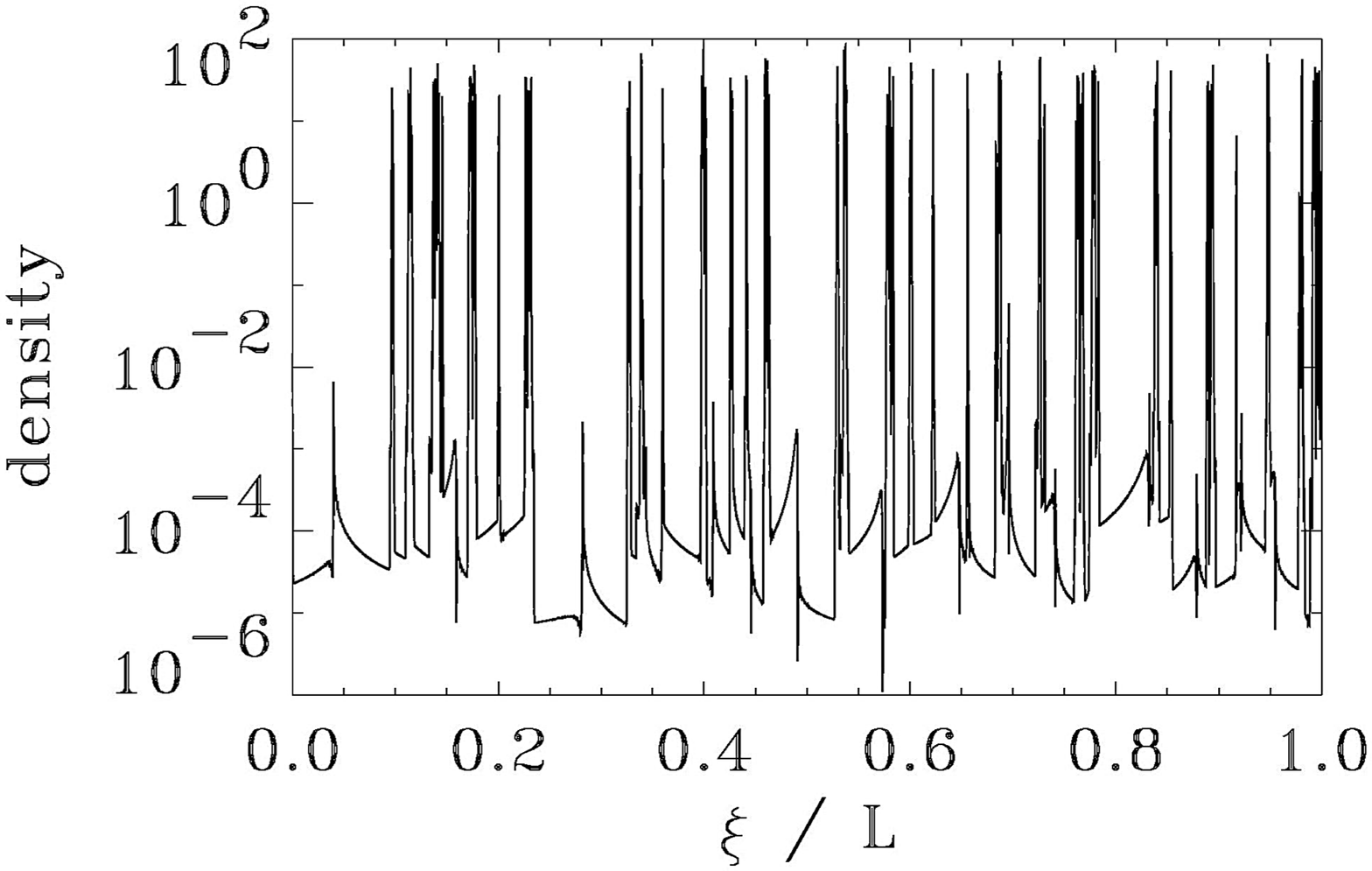}}
\caption{Matter densities at time $\theta=325$ on a semi-logarithmic scale.}
\label{fig:dens-log}
\end{figure}

Figure \ref{fig:phasespace} shows the phase-space distribution function
and corresponding matter densities $\rho$ {\modif (normalized to unity)}
at a later time. As expected the velocity domain remains bounded, so that the $N_x \times N_v$ points that mesh the phase space are used in an optimized way.
The distribution function clearly displays a hierarchical structure at different scales, with small clusters orbiting each other to form larger clusters, which in turn also revolve around each other. This hierarchy is at the basis of the fractal structure observed with N-body simulations and discussed later in this work.
The density displays many spikes, which become narrower and higher as time elapses.
{\correc
These spikes are even more apparent on a semi-log plot of the density (Fig. \ref{fig:dens-log}).
}
This structure is similar to that observed with N-body codes.

{\correc
Note that the straight segments in the phase-space plots of Fig.
\ref{fig:phasespace} (see top left panel) all have approximately the same positive slope, and correspond to regions of low matter density (``voids").
Similar behavior was seen in N-body simulations and it represents regions that are devoid of particle crossings \cite{MRexp,MRGexp,miller2010ewald}.
The slope of these segments can be estimated using Eq. \eqref{eqmotion_newscaling}, where we neglect the gravitational field $\mathcal{E}$ because in the relevant regions the density is low. We seek for a solution of the type
$\xi(\theta) = (1+\theta/3)^\gamma$. Substituting this expression into Eq. \eqref{eqmotion_newscaling}, we find that the parameter $\gamma$ must satisfy the algebraic equation:
\be
\gamma^2 + (3K-1)\gamma -9 = 0,
\label{eq:charact}
\ee
where $3K-1=3/\sqrt{2}$. Equation \eqref{eq:charact} has the positive root $\gamma=3/\sqrt{2}$ (the other root is negative and the corresponding solution is quickly damped away). Then the ratio between the phase space variables $\eta \equiv d\xi/d\theta$ and $\xi$ becomes:
\be
\frac{\eta}{\xi} = \frac{\gamma}{3+\theta}.
\label{eq:slope}
\ee
For $\theta=325$, we obtain a slope $\eta/\xi \approx 0.0065$. This is very close to the slope observed in Fig. \ref{fig:phasespace}, as can be deduced for instance from the segment on the left of the top left panel. We also checked that, for large times, the observed slope decreases as $1/\theta$, in accordance with Eq. \eqref{eq:slope}.
}

\subsection{Power spectrum}
\begin{figure}[]
\includegraphics[scale=0.4]{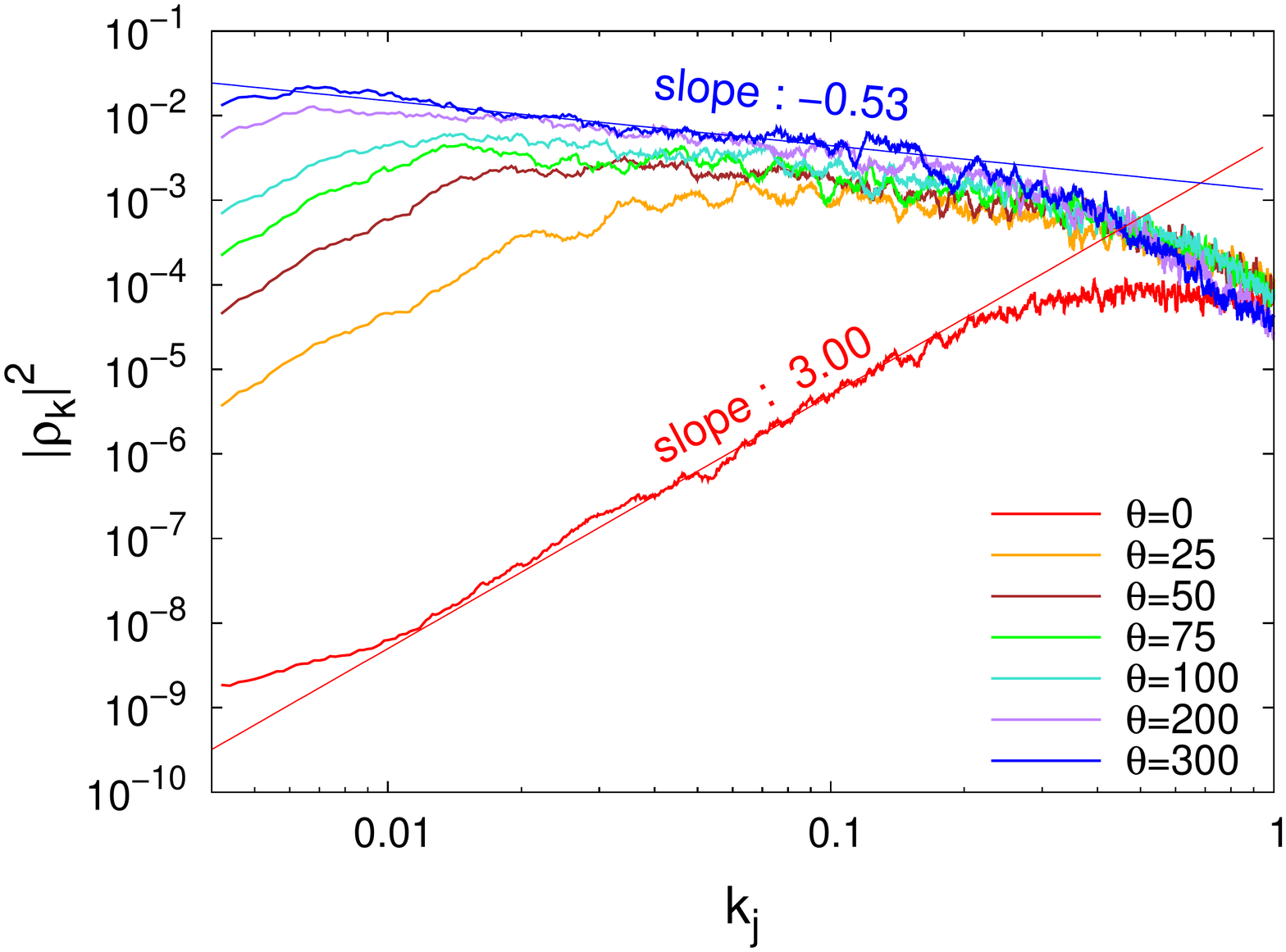}
\caption{{\it (Color online)}. Power spectrum of the matter density $|\rho_k|^2$ for different times from $\theta=0$ to $\theta=300$. Later times correspond to larger values of $|\rho_k|^2$. A moving average over 41 points is taken in order to smooth the fluctuations.}
\label{fig:spectrum}
\end{figure}

The development of hierarchical (scaling) behavior can be tracked by the evolution of the power spectra.
Here we show in Fig. \ref{fig:spectrum} the power spectrum of the matter density $|\rho_k|^2$ as a function of the wave number $k_j = (2\pi/L)j \in [2\times 10^{-4}, 3.28]$, with $j=1\dots N/2$.
Rather quickly, a decreasing power-law spectrum builds up, with a slope roughly equal to $-0.53$, not far from the value $-0.45$ observed for N-body simulations of the RF model with the same initial spectrum \cite{MRS:2015arX}.
The range of the power-law region $(k_{min},k_{max})$ increases with time, with $k_{min}$ getting smaller and smaller while $k_{max}$ remains roughly constant.
The steep decrease at $k>k_{max}$ is due to numerical diffusion.
The observed slope is also consistent with recent predictions \cite{Joyce2011,Benhaiem11032013} which, when applied to our simulations, yield a slope of $-0.57$ \footnote{Eq. (18) in Ref. \cite{Benhaiem11032013}, with $n=3$ and $\kappa=\sqrt{3}$}. Benhaiem et al. \cite{Benhaiem21092014}
found that this result also holds for 3D cosmology in scale-free models.

\begin{figure}[]
\includegraphics[scale=0.4]{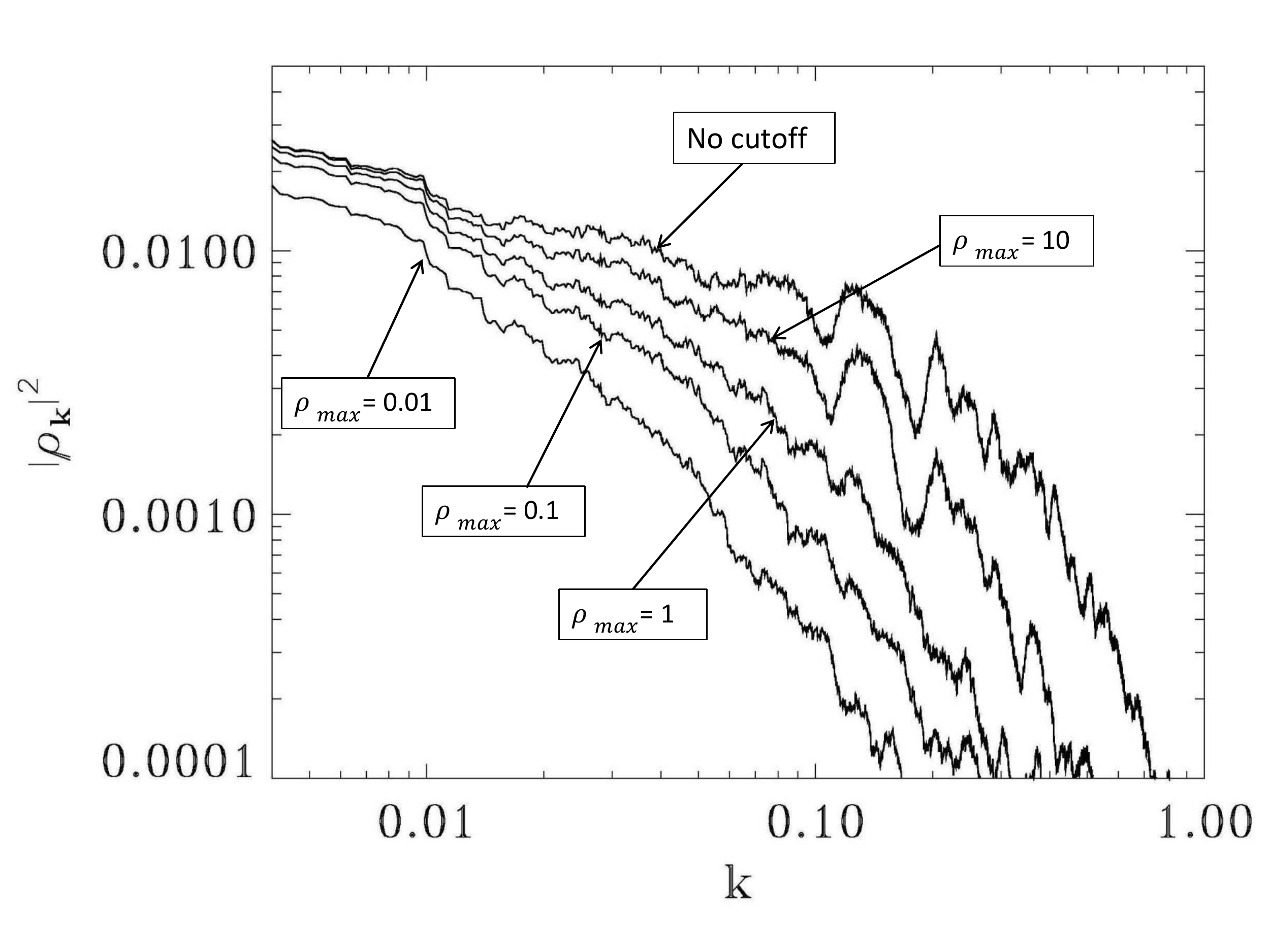}
\caption{Low-density power spectrum at $\theta=300$. For each curve, density values {\em above} the corresponding $\rho_{max}$ have been removed before computing the spectrum.}
\label{fig:spectrum_cutoff}
\end{figure}

{\correc
The power spectrum is a useful indicator of the difference between the Vlasov and N-body results. As already noted, the Vlasov power spectrum (Fig. \ref{fig:spectrum}) has a slope similar to that observed in the N-body case.
To gain further insight, we re-analyze the spectrum by performing different cut-offs, either removing the low or high values of the matter density.

Let us first consider the high-density power spectrum.
In Fig. \ref{fig:spectrum_rhomin}, we show the spectrum obtained by considering only the values of the density that are {\em above} a certain $\rho_{min}$ (values that are below this threshold are removed). We observe that the slope becomes less steep with increasing cut-off, and is almost flat for $\rho_{min}=20$. Of course, this is a huge cut-off, and we chose to show these values only to point out the general trend.
Nevertheless, this observation may explain why the observed N-body spectrum is slightly less steep ($\rm slope\approx-0.45$) than the corresponding full Vlasov result ($\rm slope\approx-0.53$): the N-body spectrum lacks the contribution from the low-density regions, which tend to steepen the spectrum as seen in Fig. \ref{fig:spectrum_rhomin}.
Consistently with this reasoning, the Vlasov spectrum (which includes both high and low densities) is closer to the analytical estimate of Ref. \cite{Benhaiem11032013} ($\rm slope\approx-0.57$).
It is also interesting to estimate the value of the cut-off that yields a slope similar to that observed in the N-body results, i.e., $-0.45$. We found that the required cut-off is $\rho_{min}\approx 4$. Notice that this is a relatively small threshold in our units, as $94.5\%$ of the matter density is above that value.

\begin{figure}[]
\includegraphics[scale=0.5,angle=90]{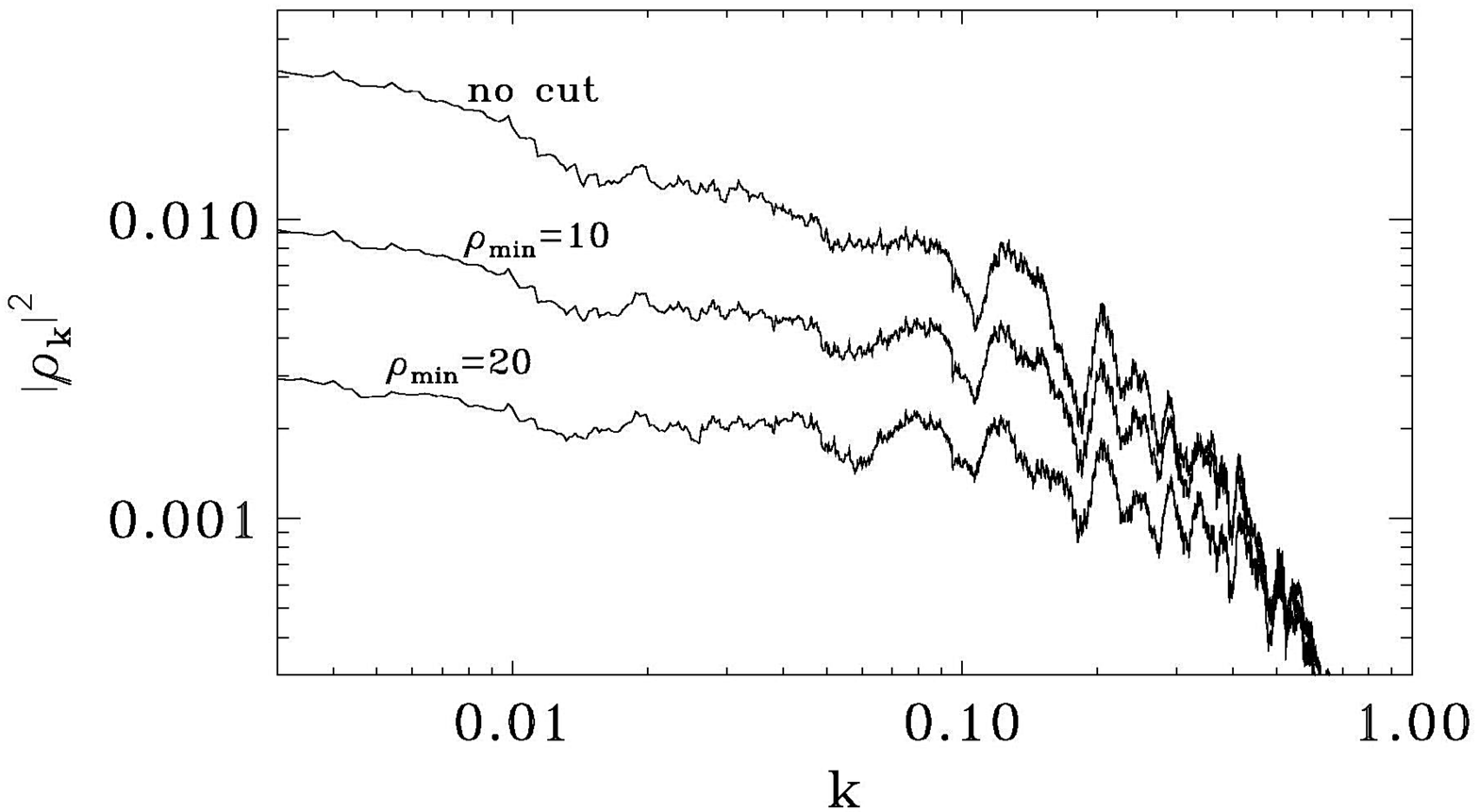}
\caption{High-density power spectrum at $\theta=300$. For each curve, density values {\em below} the corresponding $\rho_{min}$ have been removed before computing the spectrum.}
\label{fig:spectrum_rhomin}
\end{figure}

Conversely, if we consider only the low values of the density, the observed spectrum is significantly steeper, as is shown in Fig. \ref{fig:spectrum_cutoff}.
All in all, it appears that the total spectrum results from the combination of a steeper (for low densities) and a flatter (for high densities) curve. N-body codes only capture high-density regions, and therefore yield a spectrum with a slope slightly smaller than the theoretical estimate.
}

These results may be useful, for instance, to improve our understanding of the geometry and distribution of particles in voids, which is a cosmological problem of current interest \cite{Ricciardelli11092013, Szapudi11062015}.

\subsection{Fractal dimension}
The clustering of the phase-space (Fig. \ref{fig:phasespace}) and the power law
observed in the density spectrum (Fig. \ref{fig:spectrum}) point to an underlying fractal structure of the matter distribution, as was the case for the N-body simulations \cite{MRexp}.
Box counting is the method most frequently used to analyze the properties of a fractal structure \cite{Falconer}. Here, this method is used to determine the
generalized fractal dimension $D_q$ in $\xi$-space, also known as the Renyi dimension.
The system ($0\le \xi \le L$) is covered with boxes of length $\ell$ of decreasing size: $\ell=L/2$, $\ell=L/4$, and so on. The fractal dimension is defined as
\begin{eqnarray}
D_q &=& \frac{1}{q-1} \lim_{\ell \to 0}\frac{\log(\sum_i m_i^q)}{\log(\ell)}, \,\, {\rm for}\,\, q\neq 1 \label{Dq}\\
D_1 &=& \lim_{\ell \to 0}\frac{\sum_i m_i \log(m_i)}{\log(\ell)} , \,\, {\rm for} \,\, q = 1,
\end{eqnarray}
where $m_i(\ell) = \int_{i\ell}^{(i+1)\ell} \rho(\xi) d\xi/m_{tot}$ represents the proportion of mass contained in the $i$-th box, $m_{tot}$ is the total mass, and $q$ is an exponent that is intended to give more weight to either high density (when $q>0$) or low density regions ($q<0$).
To improve the statistics, the result is averaged over 1024 realizations obtained by shifting the origin of the system by multiples of the grid spacing and taking into account the periodicity of the boundary conditions.

\begin{figure}[]
\includegraphics[scale=0.4]{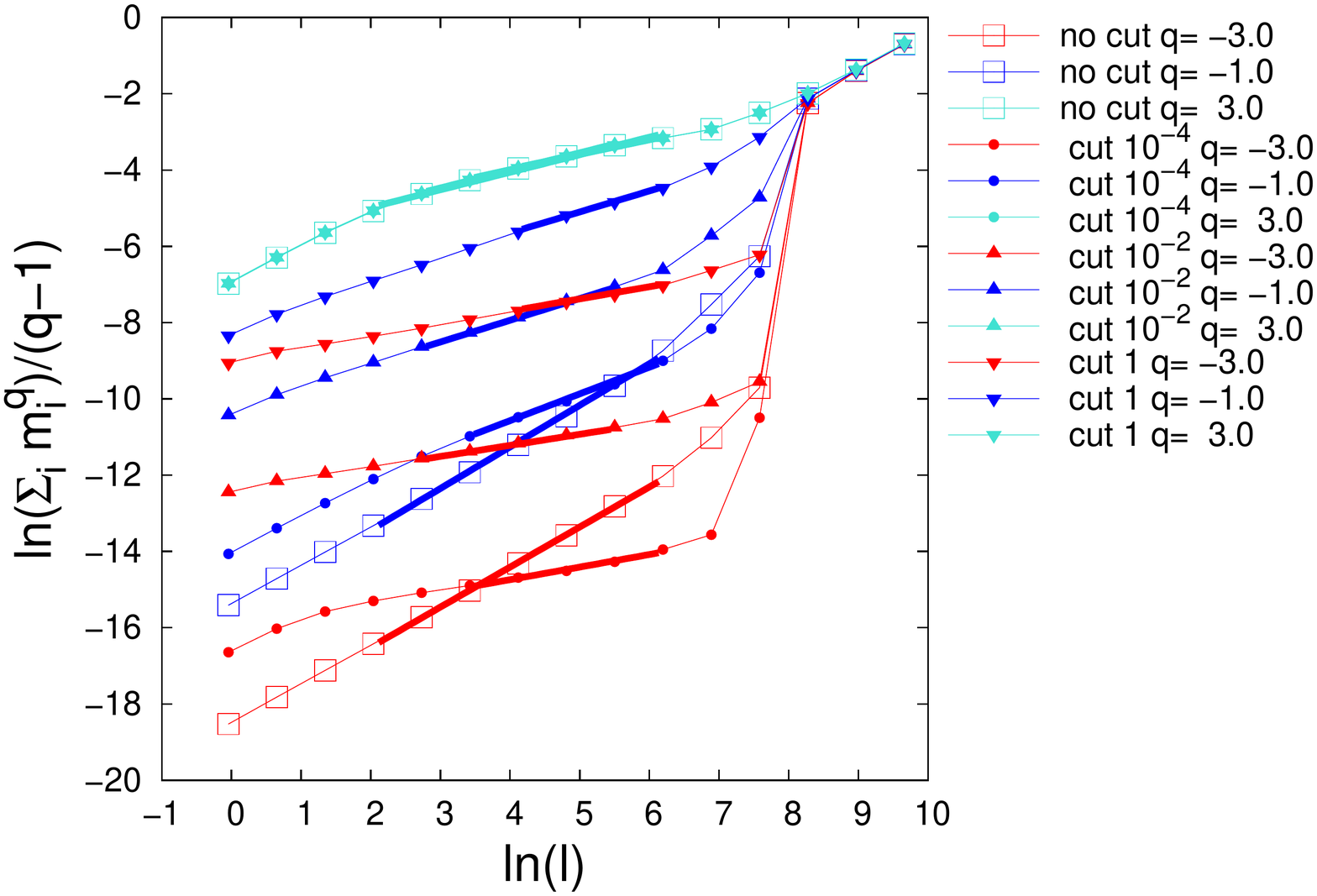}
\caption{{\it (Color online)}. Computation of the fractal dimension $D_q$ from Eq. \eqref{Dq}, for three values of $q=-3, -1$ and 3, and several cut-offs: $\rho_{th} = 10^{-4}$, $10^{-2}$, and 1. The thick lines indicate the ranges over which the slope was computed (these ranges are the same for different values of $q$ at a given cut-off).}
\label{fig:Cq}
\end{figure}

\begin{figure}[]
\includegraphics[scale=0.4]{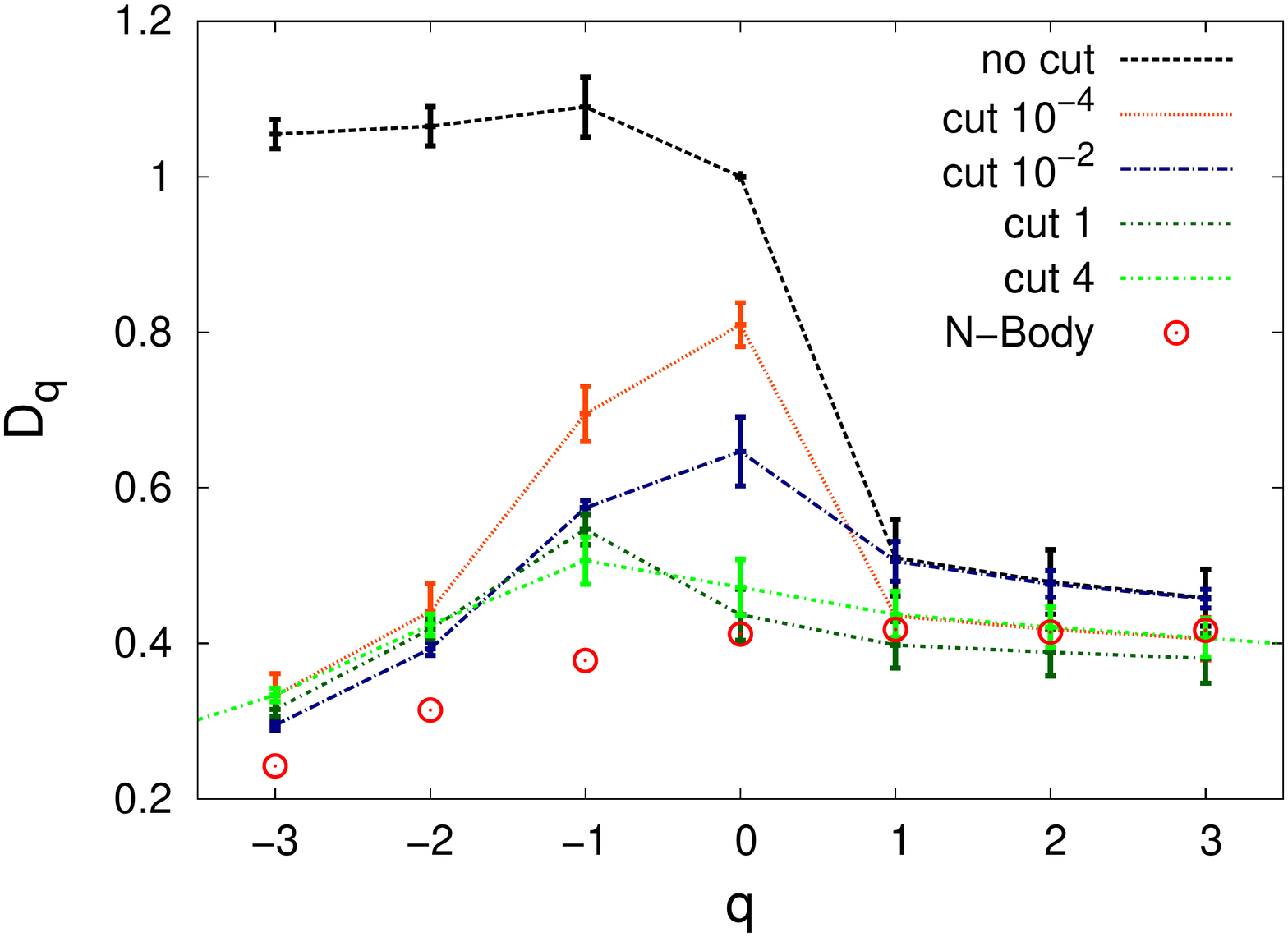}
\caption{{\it (Color online)}. Fractal dimension $D_q$ for various values of the exponent $q$ and different cut-offs of the matter density. Open circles correspond to the N-body results obtained with $N \approx 262~000$ particles. The black dashed curve corresponds to the full Vlasov results (no cut-off). Other lines correspond to Vlasov results with cut-off threshold at $\rho_{min} = 10^{-4}$, $10^{-2}$, 1, and 4.}
\label{fig:Dq}
\end{figure}

{\modif
A few examples of computation of $D_q$ are shown in Fig. \ref{fig:Cq}. In practice, $D_q$ is given by the slope of the curves for intermediate length scales. Note that for large $\ln(\ell)$ the slope is always equal to unity, signalling that the system becomes homogeneous at large scales. The uncertainty in the linear regression procedure yields the error bars that appear in  Fig. \ref{fig:Dq}.
}

It can be proven \cite{Ott} that $D_q$ should be a monotonically decreasing (or flat) function of the exponent $q$. However, N-body simulations showed that, while $D_q$ displays the expected trend for $q>0$, it is an increasing function for $q<0$ (open circles in Fig. \ref{fig:Dq}).
Since negative values of $q$ overrepresent low-density regions, this behavior was attributed to poor sampling of these regions, where the number of particles is small and the statistics noisy.

Vlasov codes, by sampling the entire phase space with the same accuracy irrespective of the matter content, should provide better results precisely in such low-density regions.
This is indeed what we observe on Fig. \ref{fig:Dq}: For positive values of $q$, which are dominated by large-density regions, the Vlasov and N-body results are in agreement; in contrast, for negative $q$ the Vlasov results (open squares)
level off at $D_q \approx 1$
\footnote{Note that the $q=0$ case is somewhat special.
For a continuous density distribution, one always has $D_0$=1, as can be checked directly from Eq. \eqref{Dq}. In contrast, for the N-body simulations, there can be regions where no particles are present, so that $D_0<1$. In the Vlasov case, when the regions of low density are ignored, then $D_0$ is no more equal to unity, as expected.}.
At face value, these findings suggest that the matter distribution is fractal at high densities (because $D_q <1$ for $q>0$), whereas it is homogeneous at low densities ($D_q \approx 1$ for $q<0$). If confirmed, this would be an important result for our understanding of the distribution of matter in the universe.
{\correc Nevertheless, it cannot be ruled out that the leveling off of $D_q$ is partly due to
numerical diffusion, which washes out the small-scale structures that develop over time.}

To understand why N-body codes fail to reproduce correctly the $q<0$ region, we  introduced an artificial cut-off in the density $\rho$ issued from the Vlasov simulations. Thus, values of $\rho$ that are below a certain threshold are set to zero. This is the same procedure that was applied earlier to the power spectrum.
In Fig. \ref{fig:Dq}, we show the results for four values of the threshold, $\rho_{min} = 10^{-4}$, $10^{-2}$, 1, and 4 {\modif (note that, although $\rho$ is normalized to unity, the fluctuations can be much larger, as seen in Fig. \ref{fig:phasespace})}. It is clear that, by increasing the cut-off level, the Vlasov results progressively move towards the N-body results.
{\correc Interestingly, we observe that the Vlasov and N-body results start to coincide for a threshold value $\rho_{th} \approx 4$. This is in agreement with the cut-off value of the density that is necessary to recover the slope of the power spectrum observed in the N-body code (see Fig. \ref{fig:spectrum_rhomin} and related discussion).
}
These findings strongly suggest that the incorrect behavior of $D_q$ observed in N-body simulations is indeed due to poor sampling of the low-density regions, and that this drawback can be overcome by using a numerical approach based on a uniform meshing of the phase space.

\section{Discussion and conclusions}\label{sec:discussion}
Most numerical simulations of self-gravitating systems are performed using N-body codes, which solve the equations of motion of a large number of interacting particles. This is an approximation, since the number of ``particles" in a real system is virtually infinite, whereas simulations are limited to a few million particles. Ideally, one should instead solve the Vlasov-Poisson equations, but this is more costly in terms of memory storage and computing time. Except for some simplified cases \cite{Yoshikawa,Hahn}, Vlasov simulations are out of reach of current computer capabilities for 3D problems, although they are now feasible for 1D problems. Here, we have shown an application of Vlasov codes to cosmological simulations of an expanding universe. A key point was the choice of the most suitable scaling factors, which map the original phase space $(x,v)$ onto a scaled phase space $(\xi,\eta)$ that is bounded for all times,  thus optimizing the number of mesh points.

The results confirmed the appearance of self-similar clustering in the phase space and a power-law spectrum similar to that observed for N-body simulations. The  box-counting fractal dimension $D_q$
is flat and close to unity for $q<0$ and decreasing for $q>0$, suggesting that the matter distribution is not the same at low and high densities.
This different behavior was also visible in the power spectra
observed at low and high densities, and may offer an insight into our understanding of large cosmological structures such as voids.
{\correc
Nevertheless, these preliminary results, which are potentially sensitive to the numerical resolution of the code, would require more studies to be fully confirmed.
}
Other approaches based on equal mass partitions \cite{yui} may provide further information on the low-density regions.

\appendix

\section{New scaling}\label{app:newscaling}
It was observed in N-body numerical simulations that the variance of the scaled velocity (``thermal" velocity) $\eta_{th}\equiv \sqrt{ \langle \eta^2 \rangle}$ grows exponentially in time. This is a problem for grid-based Vlasov simulations, since one would need to mesh a very large velocity space in order to keep the distribution function inside the computational box for all times. Therefore, we want to look for a new scaling for which the scaled velocity is bound in time, i.e., $\langle \eta^2 \rangle \sim \rm const$.

\begin{figure}[]
\includegraphics[scale=0.4]{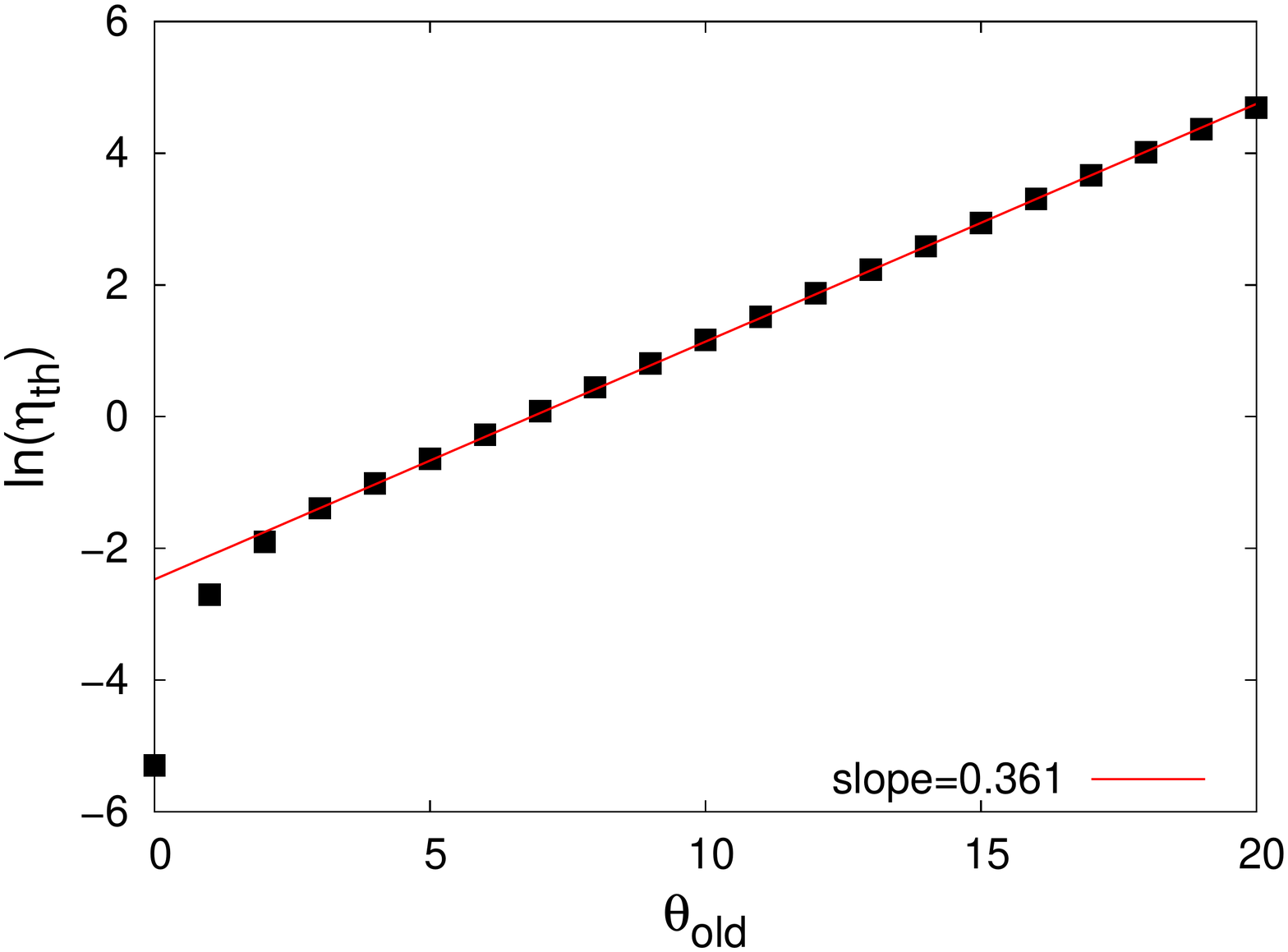}
\caption{Evolution of the scaled thermal velocity $\eta_{th}$ as a function of the scaled time $\theta_{old}$ normalized to $\omega_J^{-1}$ (standard scaling with $\beta=1$) for an N-body simulation with initial condition similar to that used for the Vlasov case. The measured slope is 0.361, close to the theoretical value 1/3 (straight line).}
\label{fig:vth}
\end{figure}

More precisely, N-body numerical simulations show that (see Fig. \ref{fig:vth}):
\be
\eta_{th}=\eta_0\,\exp\left(\frac{1}{3}\omega_{J}\,\theta_{old}\right),
\ee
{\modif
where $\theta_{old}$ is the scaled time obtained with the standard scaling ($\beta=1$).}
Using the time $t$ and remembering that $\alpha \omega_J t_0 =1$, we obtain
\be
\eta_{th}=\eta_0\,\left(\frac{t}{t_0}\right)^{1/3\alpha},
\label{etath}
\ee
The relationship between the thermal velocities $v_{th}$ and $\eta_{th}$ is deduced from Eq. (\ref{scalev}), where we neglect the last term because $\dot C$ decreases to zero with time: $v_{th}= (C/A^2) \eta_{th}$. Therefore, using Eq. (\ref{etath}), we obtain:
\be
v_{th} = \frac{C}{A^2}~ \eta_{th} =\left(\frac{t}{t_0}\right)^{-1/3} \eta_0 \left(\frac{t}{t_0}\right)^{1/3\alpha} = \eta_0 \left(\frac{t}{t_0}\right)^{(1-\alpha)/3\alpha},
\ee
where we have used the standard scaling exponents $\beta=1$ and $\gamma=2/3$.

Now we want to find a new scaling where $\eta_{th}$ is bounded in time. In Eq. (\ref{ACgeneral}), we still keep the exponent $\gamma=2/3$ (because it represents the physical expansion rate of an Einstein-de Sitter universe), and look for an exponent $\beta$ that yields $\eta_{th}\sim \rm const$.
We have:
\be
\eta_{th}= \frac{A^2}{C}~ v_{th} = \left(\frac{t}{t_0}\right)^{\beta-(2/3)}
\eta_0 \left(\frac{t}{t_0}\right)^{(1-\alpha)/3\alpha}.
\ee
In order for $\eta_{th}$ to be constant in time, one needs to satisfy:
\be
\frac{1-\alpha}{3\alpha}=\frac{2}{3}- \beta,
\ee
which yields
\be
\beta=\frac{3\alpha-1}{3\alpha}=\frac{9-\sqrt{2}}{9}\approx 0.84.
\label{beta}
\ee
This is the value of $\beta$ used in our simulations.

The new scaling derived above
\be
A^2(t) = \left(\frac{t}{t_0}\right)^\frac{9-\sqrt{2}}{9} ~,~~~~
C(t)= \left(\frac{t}{t_0}\right)^{2/3},
\label{ACnew}
\ee
yields a different equation of motion, where some of the coefficients are time-dependent.
Substituting Eq. (\ref{ACnew}) into the general form of the equation of motion (\ref{general_scaling}), one obtains
\be
\frac{\d^2 \xi}{\d \theta^2}+\frac{4-3\beta}{3\,t_0}\,\left(\frac{t}{t_0}\right)^{\beta-1}\,\frac{\d \xi}{\d \theta}-\frac{2}{9\,t_0^2}\left(\frac{t}{t_0}\right)^{2(\beta-1)}\xi=
\left(\frac{t}{t_0}\right)^{2(\beta-1)}\mathcal{E},
\label{motionbeta}
\ee
where now $\beta=\frac{9-\sqrt{2}}{9}$.
The relationship between the times $t$ and $\theta$ is obtained by integrating Eq. (\ref{scalet}) with the condition that $\theta=0$ when $t=t_0$. This yields
\be
\frac{t}{t_0}= \left(\frac{1-\beta}{t_0}~\theta+1\right)^{1/(1-\beta)}
\label{eq20}
\ee

Substituting the above expression into Eq. (\ref{motionbeta}), we obtain
\be
\frac{\d^2 \xi}{\d \theta^2}+\frac{4-3\beta}{3\,t_0}\,\frac{1}{(1-\beta)\frac{\theta}{t_0}+1}\,\frac{\d \xi}{\d \theta}-\frac{2}{9\,t_0^2}\frac{1}{[(1-\beta)\frac{\theta}{t_0}+1]^2}\,\xi=
\frac{1}{[(1-\beta)\frac{\theta}{t_0}+1]^2}\,\mathcal{E}
\label{eq21}
\ee

Using the relation (\ref{beta}) and remembering that $\alpha\omega_{J} t_0=1$, one gets:
\be
\frac{\d^2 \xi}{\d \theta^2}+\frac{\alpha+1}{3}\,\frac{\omega_{J}}{\omega_{J_0}\theta/3+1}\,\frac{\d \xi}{\d \theta}-\frac{2}{9}\,\frac{\alpha^2\omega_{J}^2}{[\omega_{J}\theta/3+1]^2}\,\xi=
\frac{\mathcal{E}}{[\omega_{J}\theta/3+1]^2}.
\label{eq22}
\ee
Defining $\hat{\mathcal{E}}=\mathcal{E}/\omega_{J}^2$ and $\hat\theta=\omega_{J}\theta$, Eq. (\ref{eq22}) becomes
\be
\frac{\d^2 \xi}{\d \hat\theta^2}+\frac{\alpha+1}{3}\,\frac{1}{\hat\theta/3+1}\,\frac{\d \xi}{\d \hat\theta}-\frac{2}{9}\frac{\alpha^2}{[\hat\theta/3+1]^2}\,\xi=
\frac{\hat{\mathcal{E}}}{[\hat\theta/3+1]^2},
\label{eq23}
\ee
and finally, with $\alpha=3/\sqrt{2}$,
\be
\frac{\d^2 \xi}{\d \hat\theta^2}+\frac{3+\sqrt{2}}{3\sqrt{2}}\,\frac{1}{\hat\theta/3+1}\,\frac{\d \xi}{\d \hat\theta}-\frac{1}{[\hat\theta/3+1]^2}\,\xi=\frac{\hat{\mathcal{E}}}{[\hat\theta/3+1]^2},
\label{eq24}
\ee
which is identical to Eq. (8) in the main text.

\section{Relation between the old and new scaled times}\label{app:time}
Let us call $\theta_{old}$ the scaled time for which $A^2(t)=t/t_0$ (standard scaling, $\beta=1$), and $\theta_{new}$ the scaled time for which $A(t)^2=(t/t_0)^\beta$, with $\beta=\frac{9-\sqrt{2}}{9}$ (new scaling).
For the standard and new scalings, the relationships between the scaled times $\theta_{old}$ and $\theta_{new}$ and the real time $t$ read as follows:
\[
\frac{t}{t_0}=\exp\left(\frac{\theta_{old}}{t_0}\right)
\]
and
\[
\frac{t}{t_0}=\left[1+(1-\beta)\frac{\theta_{new}}{t_0}\right]^{1/(1-\beta)}
\]
We take the same $t_0$ in both cases, since it is the instant at which the real and scaled times coincide. Equating the two expressions for $t/t_0$, we find:
\be
\frac{\theta_{old}}{t_0}=
\frac{1}{1-\beta}\ln\left(1+(1-\beta)\frac{\theta_{new}}{t_0}\right),
\label{oldnewtimes}
\ee
or, normalizing the scaled times to $\omega_J$:
\be
\omega_J\theta_{old} =
\frac{1}{\alpha(1-\beta)}\ln\left[1+\alpha(1-\beta)\omega_J\theta_{new}\right],
\label{oldnewtimes1}
\ee
with $\alpha(1-\beta)=1/3$.

For instance, when $\omega_J\theta_{new}=300$, we obtain $\omega_J\theta_{old} \approx 13.85$. This value is in accordance with the simulation times used for the N-body simulations using the ``old" scaling variables.

\bibliography{miller_rouet_shiozawa_v10}

\end{document}